# Electroluminescence from multi-particle exciton complexes in transition metal dichalcogenide semiconductors

Matthias Paur,[1, ‡] Aday J. Molina-Mendoza,[1, ‡,*] Rudolf Bratschitsch,[2] Kenji Watanabe,[3] Takashi Taniguchi,[3] and Thomas Mueller[1,*]

[1] *Vienna University of Technology, Institute of Photonics, Gußhausstraße 27-29, 1040 Vienna, Austria*

[2] *Institute of Physics and Center for Nanotechnology, University of Münster, Wilhelm-Klemm-Strasse 10, 48149 Münster, Germany*

[3] *National Institute for Materials Science, 1-1 Namiki, Tsukuba, 305-0044 Japan*

‡ These authors contributed equally to this work.
* Corresponding authors: aday.molina-mendoza@tuwien.ac.at, thomas.mueller@tuwien.ac.at

**ABSTRACT**

Light emission from higher-order correlated excitonic states has been recently reported in hBN-encapsulated monolayer WSe$_2$ and WS$_2$ upon optical excitation. These exciton complexes are found to be bound states of excitons residing in opposite valleys in momentum space, a promising feature that could be employed in valleytronics or other novel optoelectronic devices. However, electrically-driven light emission from such exciton species is still lacking. Here we report electroluminescence from bright and dark excitons, negatively charged trions and neutral and negatively charged biexcitons, generated by a pulsed gate voltage, in hexagonal boron nitride encapsulated monolayer WSe$_2$ and WS$_2$ with graphene as electrode. By tailoring the pulse parameters we are able to tune the emission intensity of the different exciton species in both materials. We find the electroluminescence from charged biexcitons and dark excitons to be as narrow as 2.8 meV.

INTRODUCTION

Monolayer transition metal dichalcogenide (TMD) semiconductors provide a unique platform to study light-matter interaction and many-body effects at the atomic scale. The strong Coulomb interaction in these materials leads to the formation of tightly-bound electron-hole pairs (excitons) with binding energies of hundreds of milli-electronvolts[1–3]. Moreover, excitons residing in the two different valleys at the K points of the Brillouin zone can also interact[4–10], giving rise to four- and five-particle states, which have been recently identified in tungsten diselenide ($WSe_2$) as neutral and charged biexcitons[11–14], as well as biexcitons in $WS_2$[15,16]. The radiative emission from such exciton complexes can further be tuned by an external electric field, modifying the doping of the material and thus favoring the formation of either charged or neutral complexes. This rich and complex excitonic scenario could be exploited to develop applications such as valleytronics and (quantum) optoelectronic devices, especially in the case of electrically-driven light emitters. Electroluminescence (EL) from monolayer TMDs has been widely studied in the past, especially at room temperature, where the emission originates from neutral excitons[17–26]. At cryogenic temperatures, the emission can also be due to localized exciton states or negatively charged trions[27–33]. However, electrically driven emission from more complex exciton species has not yet been observed.

Here, we report on EL from monolayer $WSe_2$ and $WS_2$ by pulsed transient EL[34], which triggers the formation of exciton complexes and thus their light emission. The high sample quality, enabled by encapsulating the monolayer semiconductor in hexagonal boron nitride (hBN), allows for the observation of EL from multi-particle exciton complexes, including neutral and negatively charged biexcitons, with narrow emission linewidths down to ~2.8 meV (full-width-at-half-maximum (FWHM)). Furthermore, the formation of different exciton species is tunable by tailoring of the electrical pulse parameters. Our results provide crucial insights into electrically-driven light emission in monolayer TMDs and open a new route to applications in novel optoelectronic devices.

## RESULTS

### Sample structure and light emission at room temperature

The sample structure used to study EL in monolayer TMDs is schematically depicted in Fig. 1a. A microscope image of the $WS_2$ sample presented in this article is shown in Fig. 1b (a microscope image of the $WSe_2$ sample is shown in Supplementary Figure 1): the monolayer TMD (namely $WSe_2$ or $WS_2$) and graphene are encapsulated in between two multilayer hBN flakes. The graphene layer is used as electrical contact to the monolayer semiconductor. Part of the graphene flake is left un-encapsulated in order to make an electrical contact with a Pd/Au electrode. The top hBN serves as encapsulation[35] to avoid environment related effects (residues left after sample processing, adsorbents) on the light emission from TMDs, and the bottom hBN also provides an insulating layer to apply a gate voltage to the sample by means of a pre-patterned Ti/Au back-gate electrode, on top of which the entire heterostructure is placed. Graphene was chosen as electrode for two reasons: (i) it is a layered material that can be easily integrated in a van der Waals heterostructure, and (ii) it has been shown that charge carrier injection from graphene for transient EL is more efficient than from standard metals[34], resulting in enhanced emission. Details of the sample fabrication can be found in the Methods section.

This sample structure (metal-insulator-semiconductor-graphene) allows for electrically-driven light emission by the recently reported method of transient EL in transition metal dichalcogenides[34], in which light emission is achieved by alternated injection of electrons and holes into the semiconductor by means of a pulsed gate voltage. In Fig. 1c we show band diagrams that schematically depict the emission process. The shape of the (~9 ns long) electrical pulses used for EL is schematically shown in the inset of Fig. 1c. The gate voltage sweeps from a negative value ($V_0$), where the Fermi level in graphene is below the Dirac point and holes are accumulated in the valence band of the semiconductor (Fig. 1c left), to a positive value ($V_1$), where the Fermi level in graphene shifts upwards in energy and electrons are injected into the conduction band (Fig. 1c right). The steep (lateral) band bending in the TMD due to the large voltage drop at the Schottky contact leads to a transient tunneling current. As a result, injected electrons diffuse in the semiconductor while holes (radiatively) recombine with incoming electrons or partially drift out through the contact[34]. This process is known as transient EL, as the injected carriers recombine only during the rise and fall periods of the pulse. The light generation can be depicted from Fig. 1d, where we show an optical microscopy picture of the sample presented in Fig. 1b, with a clear red-orangish light emission originating from the WS$_2$ flake.

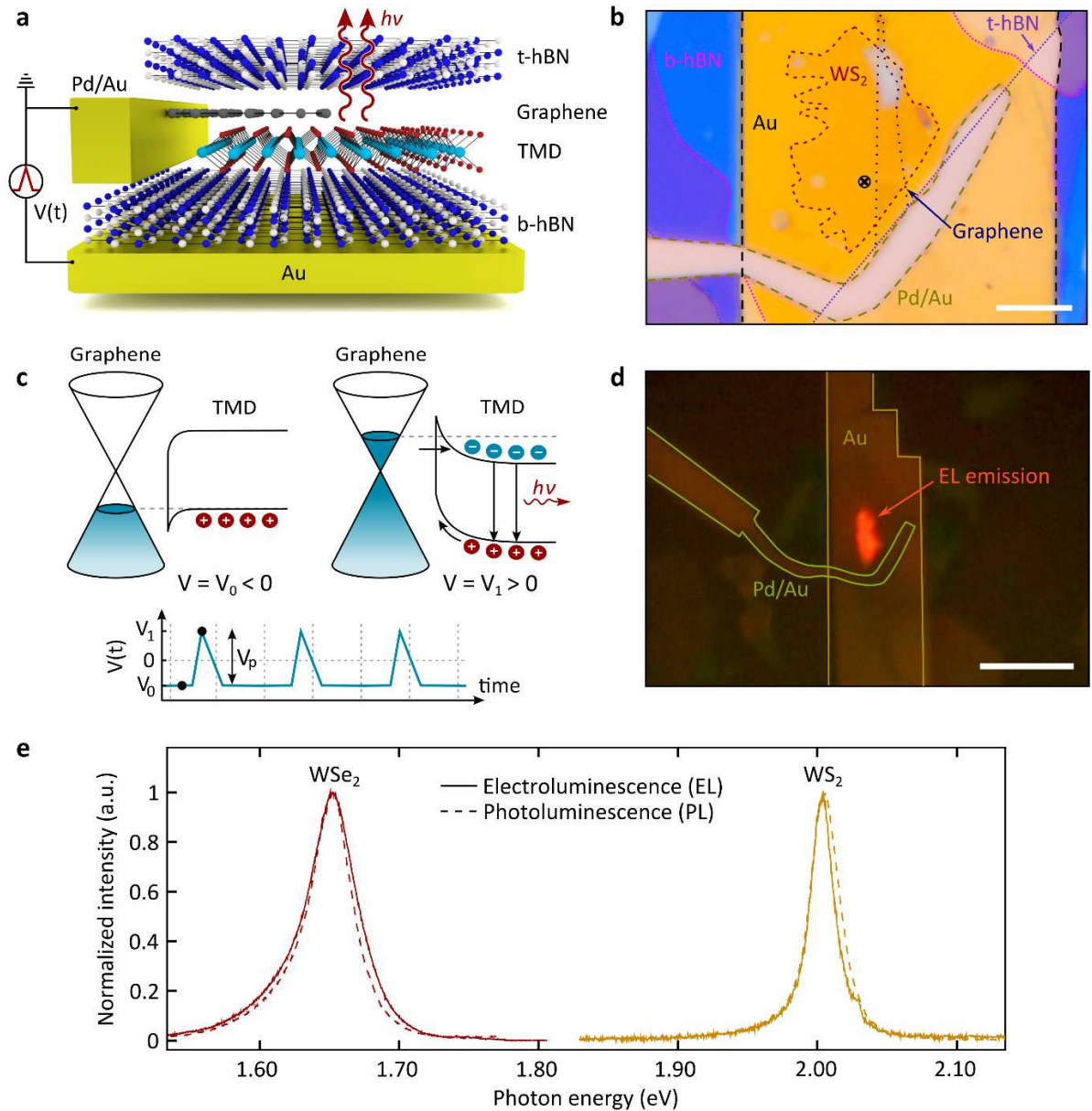

**Figure 1 | Sample structure and light emission at T = 300 K. a,** Schematic drawing of the sample. The monolayer TMD and graphene are sandwiched between two multilayer hBN flakes. **b,** Optical microscope image of a typical sample, in this case based on WS$_2$. The cross marks the position at which the PL spectra were taken. Scale bar, 5 μm. **c,** Band diagram of transient EL. On the left, the pulse voltage is at a negative value, while on the right it is at a positive value. The bottom picture schematically depicts the shape of the pulse. **d,** Optical microscopy image of the sample while emitting light. The EL emission is generated near the interface between graphene and monolayer WSe$_2$ due to carrier diffusion from graphene. Scale bar, 20

μm. **e,** Photoluminescence (PL) and electroluminescence (EL) emission spectrum (excitation $\lambda$ = 532 nm, $P_d$ = 2×10³ Wcm⁻²). The EL spectrum closely matches the PL spectrum.

The samples are characterized at room temperature by their PL spectrum (Fig. 1e) upon optical excitation with a continuous-wave laser ($\lambda$ = 532 nm, $P_d$ = 2×10³ Wcm⁻²), where we observe an emission maximum centered at ~1.653 eV for WSe$_2$ and at ~2.003 eV in the case of WS$_2$, corresponding to their respective free exciton emission (comparison between spatial resolution of PL and EL is presented in Supplementary Figures 1a and 2). Similar spectra are observed when an AC voltage is applied between the back-gate and the graphene layer, inducing transient EL near the interface between the TMD monolayer and graphene.

**Electro- and photoluminescence from exciton complexes at *T* = 5 K**

At room temperature, only one emission maximum can be identified in the PL and EL spectra of both WSe$_2$ and WS$_2$. However, the scenario becomes more complex when the sample is studied at cryogenic temperatures. At $T$ = 5 K different emission peaks can now be distinguished in the EL spectra of both materials. In WSe$_2$ (Fig. 2a), apart from the well-known bright exciton ($X^0$) centered at 1.7315 eV, the EL spectrum reveals the emission lines associated with other exciton complexes, previously observed upon optical excitation[11–14]. On one hand, we are able to distinguish the two negatively charged trions at 1.7007 eV and 1.6921 eV, attributable to the intravalley ($X_1^-$) and intervalley trions ($X_2^-$), respectively[36], and the spin-forbidden dark exciton ($X^D$) at 1.6864 eV, which is the two-particle ground state exciton[6,37,38]. Furthermore, the neutral biexciton (XX) centered at 1.7121 eV, a four-particle complex, is formed from a long-lived dark (spin-forbidden) and a bright (spin-allowed) exciton, while the charged biexciton centered at 1.6808 eV (XX⁻), a five-particle complex, is formed from a negatively charged trion and a neutral exciton in different K and K' valleys[8,11–14]. It is worth mentioning that the pulsed EL from WSe$_2$ shows a narrow emission linewidth of ~2.8 meV FWHM only for both the charged biexciton and the dark exciton. Finally, we also observe a broad peak at ~1.675 eV which may be assigned to a positively or negatively charged dark trion ($X^{D\pm}$) [39]. At energies lower than ~1.67 eV, additional emission peaks (labelled $L_1$, $L_2$ and $L_3$) emerge that most probably stem from localized states[37,40–44]. In Table 1, we summarize the energy of the different exciton complexes with respect to the bright exciton $X^0$.

A similar scenario is uncovered in the EL spectrum of WS$_2$, shown in Fig. 2b (Table 1 provides again a summary of the exciton energies). The bright exciton $X^0$ appears at an energy of 2.075 eV, as well as the charged trions $X_1^-$ and $X_2^-$ at 2.045 eV and 2.038 eV, respectively. The peak appearing at 2.023 eV has previously been associated to a biexciton[15,16], although its charged or neutral nature remains still unclear. Here we anticipate that it is due to a negatively charged biexciton (XX$^-$), since its peak height (emission intensity hereafter) increases when a positive gate voltage is applied to the sample, as will be discussed later. Furthermore, we are able to observe additional emission lines corresponding to a dark exciton ($X^D$), centered at 2.028 eV (only observed so far in the PL of non-encapsulated samples[45]) and to a neutral biexciton (XX), centered at 2.056 eV, which, to our knowledge, has not been previously reported. Importantly, the latter two exciton complexes are hardly visible in the PL spectrum, but can be clearly resolved in the respective EL. To our knowledge, this is the first evidence of EL of such multi-particle complexes in both WSe$_2$ and WS$_2$.

|  | $X^0$ (eV) | $X^0$ - XX (meV) | $X^0$ - $X_1^-$ (meV) | $X^0$ - $X_2^-$ (meV) | $X^0$ - $X^D$ (meV) | $X^0$ - XX$^-$ (meV) |
|---|---|---|---|---|---|---|
| **WSe$_2$** | 1.7315 | 19.4 | 30.8 | 39.4 | 45.1 | 50.7 |
| **WS$_2$** | 2.0752 | 19.2 | 30.2 | 36.6 | 46.36 | 52.4 |

**Table 1.** Energy of the different exciton complexes with respect to the bright exciton $X^0$ appearing in the EL spectra.

The rich EL scenario in our samples, based on charged and neutral excitonic complexes, motivates the operation of such heterostructures as tunable light emitters by just tailoring the pulse parameters. By shifting the offset of the pulsed voltage, it is possible to create either an electron-rich or a hole-rich environment in the 2D semiconductor, consequently enhancing or diminishing EL from the different exciton complexes in a similar way as it is commonly observed in gated PL measurements. To this end, according to the schematic pulse diagram shown in Fig. 1c, the EL experiment is performed as follows (see also Supplementary Figure 3): the pulse amplitude $V_p = |V_1 - V_0|$ is kept constant in all measurements, while the offset voltage $V_0$ is varied over a certain range. In this situation, an offset of $V_0 = 0$ indicates that the voltage applied to the back-gate sweeps in a pulse cycle from 0 to $V_p$, while with an offset of $V_0 = -V_p$, the voltage applied to the back-gate sweeps from $-V_p$ to 0. Therefore, in the former case an electron-rich ($p^-/n^+$) environment is induced in the semiconductor, while in the latter the injected charge carrier density will be hole-rich ($p^+/n^-$). Finally, in the symmetric case, $V_0 = -V_p/2$, the amount of injected electrons and holes is balanced (p/n). Note that this picture applies to the case of undoped devices with contacts that are equally transparent for both electrons and holes. In realistic samples the different regimes may occur at other offset voltages, but the qualitative picture remains the same.

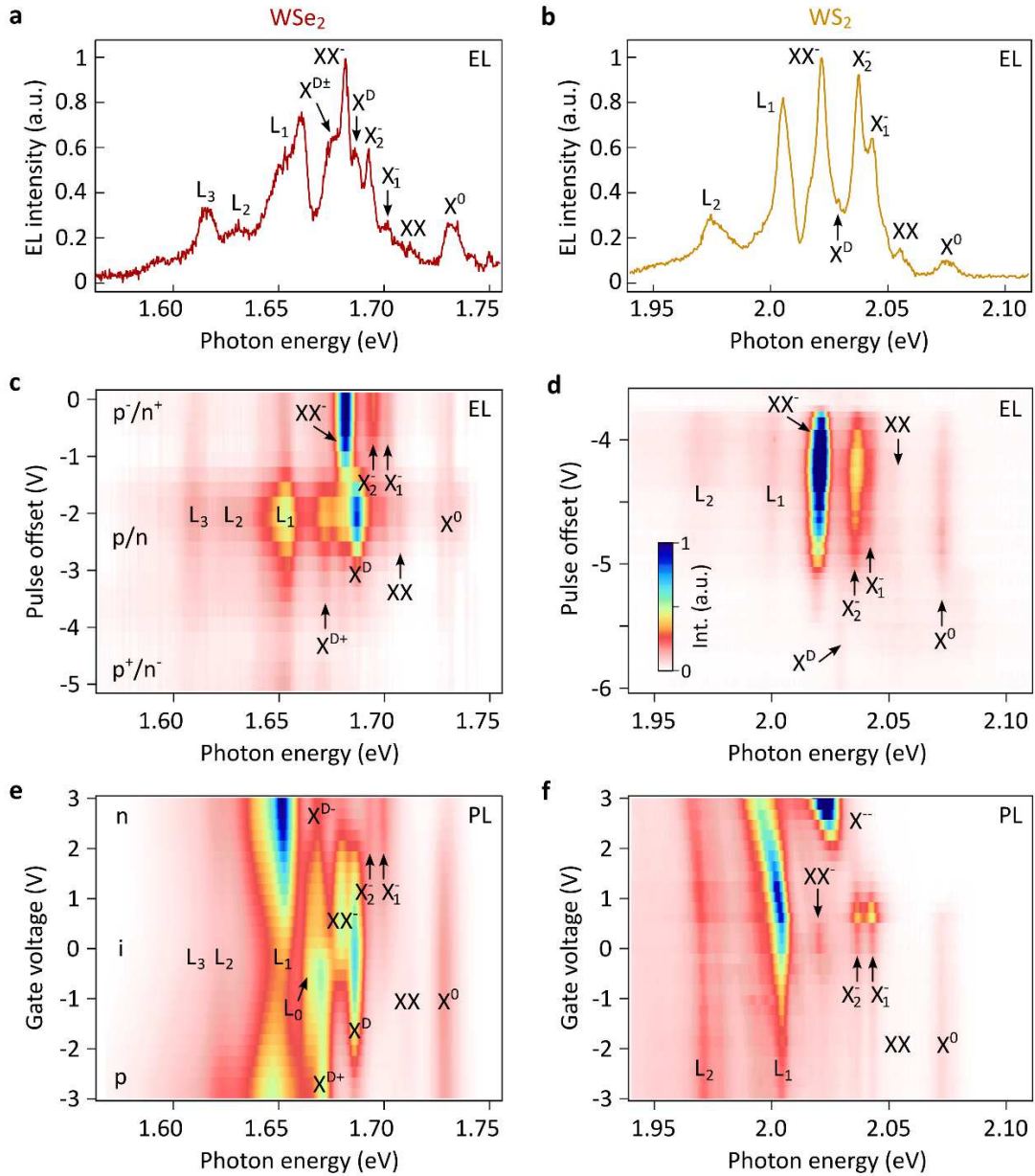

**Figure 2 | Emission spectra and tunability at $T$ = 5 K.** Pulsed EL from **a,** WSe$_2$ and **b,** WS$_2$. In WSe$_2$, a narrow emission (linewidth ~2.8 meV FWHM) is found for both the charged biexciton and the dark exciton. Colormap of EL spectra at 5 K plotted as function of pulse offset $V_0$ in **c,** WSe$_2$ and **d,** WS$_2$ ($V_p$ = 7.5 V). Colormap of charged doping control of PL emission in **e,** WSe$_2$ and **f,** WS$_2$, for different exciton species acquired at an excitation of $P_d$ = 5×10³ Wcm⁻².

In Figs. 2c and 2d we show colormaps representing the EL emission from WSe$_2$ and WS$_2$, respectively, for different $V_0$ values and with fixed pulse amplitude of $V_p$ = 7.0 V. We observe in both materials that for higher offset values (p⁻/n⁺ region), i.e. higher electron injection, the EL is dominated by negatively charged exciton species such as XX⁻, $X_1^-$ and $X_2^-$, where the charged biexciton is the predominant emission and the free exciton is much weaker. For lower offset values (p/n region), the emission from negatively charge exciton complexes vanish and the spectrum is then dominated by neutral species such as the dark and bright excitons and neutral biexciton, as well as localized excitons. This indicates that the charge carrier injection is balanced between electrons and holes, favoring the formation of neutral excitons. Finally, when the offset voltage has a more negative value (p⁺/n⁻ region), the induced charge density in the semiconductor is then hole-rich and the emission from positively charged species such as the positive dark trion in WSe$_2$ becomes stronger. We are not able to resolve positively charged trions in EL neither in WSe$_2$ or WS$_2$, possibly because the induced hole concentration is too low to favor trion formation.

The tunable EL can be compared to the respective PL colormaps (Figs. 2e and 2f), where the PL emission is measured for different DC gate voltages ($V_g$). In a similar manner as for EL, negative exciton species become the dominant emission for positive applied gate voltages ($V_g > 0$), i.e. for n-doping, although in the PL spectra we are able to additionally resolve the negative dark trion in WSe$_2$ and the double-negatively charged trion (X⁻⁻), the next charging state of the trion[13], in WS$_2$. In the WS$_2$ spectra we also observe XX⁻ emission only in the n-doped region, supporting the negatively charged nature of this biexciton as concluded above from the EL measurements. Around $V_g = 0$, the negatively charged species vanish and only the neutral excitons and biexciton, as well as the localized states, show emission, again in accordance with the EL spectra. Here it is worth noting that we are not aware of any previous reports on the PL emission of neutral biexcitons in WS$_2$, which we are able to resolve both in EL and PL. Finally, for $V_g < 0$, the positive dark trion appears in the WSe$_2$ PL spectrum, but not in the one of WS$_2$, suggesting that WS$_2$ is unintentionally n-doped and we therefore cannot achieve sufficient p-doping in this material. We refer the reader to the Supplementary Figures 4 and 5 for a further analysis of the PL spectra as a function of the back-gate voltage.

**Discrimination between two-, three-, four- and five-particle excitons**

The electrical tunability of the EL and PL spectra permits the discrimination between charged and neutral exciton species. Whether they are formed by two, three, four or five particles cannot be distinguished by this method. We therefore studied the dependence of the emission intensity with the excitation intensity. While the emission intensities of dark and bright excitons, composed by one electron and one hole, follow a linear dependence with the excitation power, biexcitons are expected to show a quadratic behavior[11–16]. In Supplementary Figures 6, 7 and 8 we show the PL spectra of $WSe_2$ and $WS_2$, at different excitation powers and fits of the respective intensities to the power-law $I_{PL} \propto P^\alpha$, where $I_{PL}$ is the integrated PL emission and $P$ the incident laser power. These fits yield the exponent values $\alpha$ = 1.01 for $X^0$, $\alpha$ = 1.82 for XX, and $\alpha$ = 1.44 for XX⁻ in $WSe_2$, and similar values in $WS_2$ (see Supplementary Table 1). The deviation from the quadratic growth for the charged biexciton can be attributed to the lack of equilibrium between exitonic states[46,47].

We have also studied the dependence of the emission intensity on the excitation power in EL by keeping a fixed offset voltage and changing the electrical pulse amplitude. For increasing pulse amplitude, the Fermi level shifts deeper in the conduction band and, therefore, more charge carriers are injected, resulting in increased EL emission. The measured spectra are shown in Figs. 3a and 3c for $WSe_2$ and $WS_2$, respectively, and the corresponding plots of EL intensities against pulse amplitude are presented in Figs. 3b and 3d. On a semi-logarithmic scale, the dependence of EL intensity versus voltage shows a linear behavior for all exciton species. This observation motivates us to empirically model the EL intensity by the expression $I_{EL} \propto \exp(\alpha \kappa V_p)$, where $V_p$ is the applied voltage, $\kappa$ is a constant related to the injected carrier concentration in the 2D semiconductor, and $\alpha$ is a coefficient that indicates the nature of the exciton, as in the PL measurements. We fix $\kappa$ to fit the data for the neutral exciton $X^0$ with $\alpha \equiv 1$, which then serves as reference for determining the $\alpha$ values of all other exciton species.

The fit of the measurement data to this empirical expression yields for WSe$_2$ a coefficient of $\alpha =$ 1.75 for XX$^-$, which is significantly larger than that of the X$^0$ and X$^D$ excitons ($\alpha \equiv 1$ and $\alpha = 1.03$, respectively) and it is close to a quadratic dependence. For the neutral biexciton XX we obtain a coefficient of $\alpha = 1.44$. For WS$_2$, the coefficients for X$^0$, X$_1^-$ and X$_2^-$ are $\alpha \equiv 1$, $\alpha = 1.32$ and $\alpha = 1.33$, respectively, and $\alpha = 2.33$ and $\alpha = 2.31$ for the neutral biexciton XX and charged biexciton XX$^-$. In Supplementary Table 1 we summarize the obtained coefficients for the different exciton complexes. The emission of all exciton species in WS$_2$ saturate at voltage amplitudes larger than ~8 V, in accordance with measurements performed at room temperature by Lien *et al.*[34]. We do not observe such saturation in WSe$_2$, most probably because of the lower peak voltage applied in these measurements. Finally, we present EL spectra for both WSe$_2$ and WS$_2$ at different pulse repetition rates in Supplementary Figure 9.

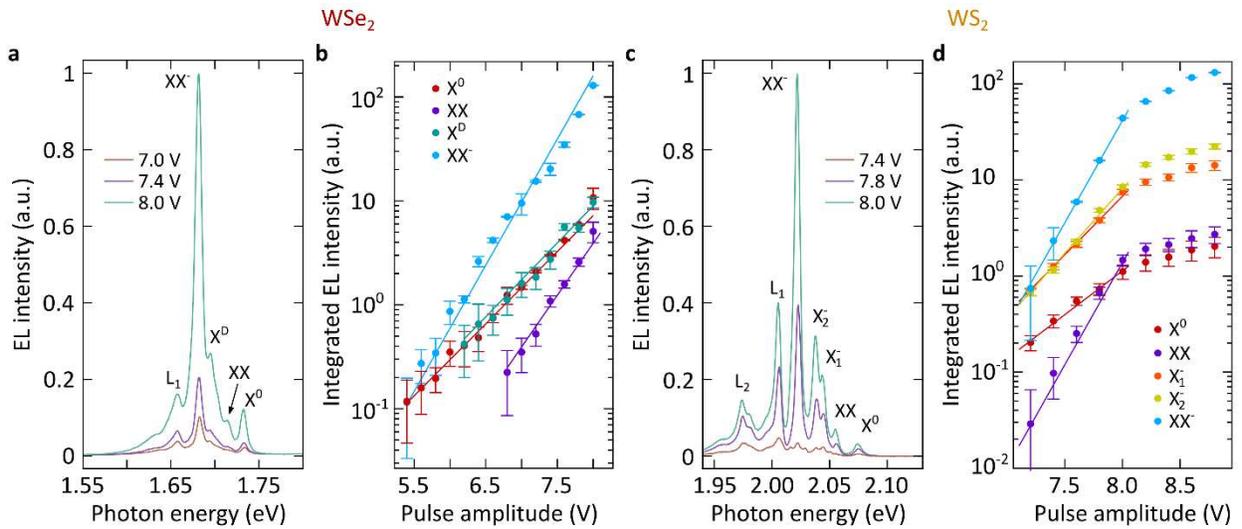

**Figure 3 | EL intensity dependence with the excitation power. a,** Pulsed EL emission spectra of WSe$_2$ for different pulse amplitudes and fixed offset voltage. **b,** Semi-logarithmic plot of the integrated EL intensity as a function of the pulse amplitude. The circles represent the experimental data, while the lines depict the fit to an exponential function as discussed in the main text. **c,** and **d,** Same as in a, and b, but for WS$_2$.

**DISCUSSION**

In summary, we have studied transient EL of monolayer TMD semiconductors (in particular WSe$_2$ and WS$_2$) sandwiched between two layers of hBN and one graphene sheet at low temperatures. The high material quality of our samples, together with the strong Coulomb interaction between electrons and holes in 2D semiconductors, permits the observation of electrically-driven light emission of higher-order correlated excitonic states, including bright and dark excitons, negatively charged trions and neutral and negatively charged biexcitons. By tailoring the pulse parameters, it is possible to create either an electron- or a hole-rich environment in the 2D semiconductor, consequently favoring the enhanced or diminished EL from the different exciton species. Our technique extends and complements gate-dependent PL spectroscopy and will enable further investigations of many-body phenomena in 2D materials. From an applied point of view, our devices may find application as wavelength tunable light emitters or furnish new opportunities for quantum light sources, e.g. by quantum-confinement of electrically induced biexcitons.

**METHODS**

**Sample fabrication.** Bulk $WSe_2$ (grown by vapor phase transport; see Ref. [48] for details) and $WS_2$ (purchased from *HQ Graphene*), graphite and hBN were exfoliated by an adhesive tape and transferred onto a $SiO_2$ substrate thermally grown on Si. The substrates were cleaned by an acetone/isopropanol bath and oxygen plasma treatment. They were then annealed at 300 °C for 2 hours in a dry-air atmosphere prior to the transfer of the exfoliated materials to remove possible adsorbates. The tape was peeled-off afterwards and the atomically-thin flakes were selected with optical microscopy. A pick-up and place technique[49] was used to pick up the flakes by means of a polypropylene carbonate/polydimethylsiloxane (PPC/PDMS) stamp, starting with the top hBN flake. The heterostructure (hBN/graphene/$WSe_2$/hBN) was then transferred onto a Ti/Au (5/100 nm) electrode, pre-patterned on a $SiO_2$/Si substrate. The entire exfoliation and transfer process was performed in a nitrogen-purged glovebox. The remaining PPC residues were removed by introducing the sample in an acetone/chloroform bath. Finally, an additional electrode was fabricated by means of electron-beam lithography and metal evaporation to make an electrical contact to the graphene flake (Pd/Au, 60/140 nm).

**PL and EL measurements.** For EL measurements a pulse generator (Agilent 8114A) in combination with a source-meter unit (Keithley 2612A) was used. The pulses were applied to the bottom electrode, while the top contact was grounded. Optical and electrical investigations were carried out in a liquid-helium exchange gas cryostat (Oxford Instruments) under high vacuum ($10^{-7}$ mbar), integrated in a scanning micro-PL setup. The setup includes a dichroic mirror at $\lambda = 550$ nm to separate the PL emission from the excitation. PL and EL were collected using a 50× long working distance objective lens (NA = 0.5). In PL experiments, the sample was optically excited non-resonantly using a continuous-wave 532 nm laser; the diameter of the laser beam on the sample was ~1 µm. The PL/EL emission was spectrally filtered with a monochromator (Horiba iHR320) and detected with a liquid nitrogen cooled charge-coupled device (CCD). EL imaging was performed by using a CCD camera (Thorlabs).

**Acknowledgments:** We acknowledge financial support by the Austrian Science Fund FWF (START Y 539-N16) and the European Union (grant agreement No. 785219 Graphene Flagship). A.J.M-M. acknowledges financial support from the European Commission (Marie Sklodowska-Curie Individual Fellowships, OPTOvanderWAALS, grant ID 791536). M.P. acknowledges financial support from the doctoral college program "TU-D" funded by TU Vienna. K.W. and T.T. acknowledge support from the Elemental Strategy Initiative conducted by the MEXT, Japan and the CREST (JPMJCR15F3), JST.

**Authors contributions:** A.J.M-M. and T.M. conceived the project. M.P. and A.J.M-M. fabricated the samples and performed the measurements. R.B. provided high-quality $WSe_2$ crystals. T.T. and K.W. synthetized the hBN crystal. M.P., A.J.M-M. and T.M. wrote the manuscript with input from all the authors. All authors discussed the results and contributed to the manuscript.

**Competing financial interests:** The authors declare no competing financial interests.

**Data availability:** The data that support the findings of this study are available from the authors upon reasonable request; see authors contributions for specific data.

*Supplementary Information*

# Electroluminescence from multi-particle exciton complexes in transition metal dichalcogenide semiconductors

**Matthias Paur,**[1, ‡] **Aday J. Molina-Mendoza,**[1, ‡,*] **Rudolf Bratschitsch,**[2] **Kenji Watanabe,**[3] **Takashi Taniguchi,**[3] **and Thomas Mueller**[1,*]

[1] *Vienna University of Technology, Institute of Photonics, Gußhausstraße 27-29, 1040 Vienna, Austria*

[2] *Institute of Physics and Center for Nanotechnology, University of Münster, Wilhelm-Klemm-Strasse 10, 48149 Münster, Germany*

[3] *National Institute for Materials Science, 1-1 Namiki, Tsukuba, 305-0044 Japan*

‡ These authors contributed equally to this work.

* Corresponding authors: aday.molina-mendoza@tuwien.ac.at, thomas.mueller@tuwien.ac.at

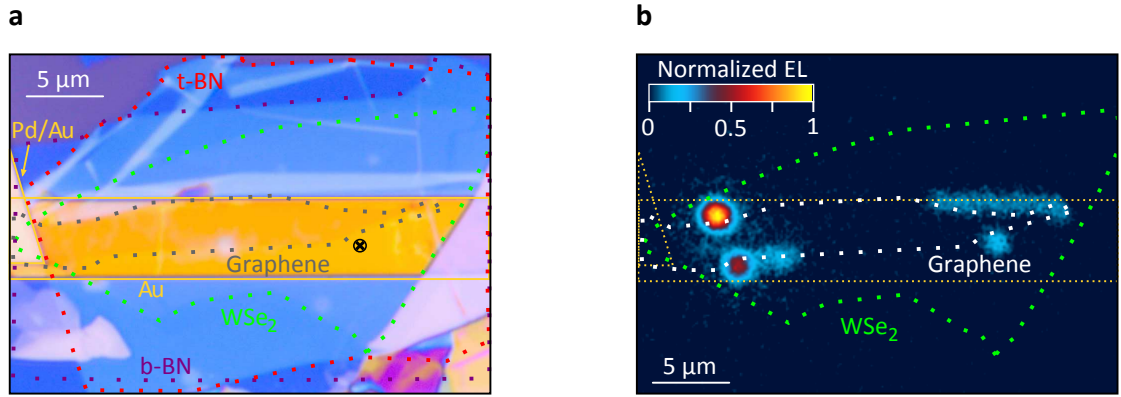

**Supplementary Figure 1 | Electroluminescence image of WSe$_2$ sample at 300 K. a,** Optical microscopy image of the sample. **b,** False-color EL image of the sample. The EL emission is generated near the interface between graphene and monolayer WSe$_2$. However, the light emission is not homogenously generated along this interface, because part of the graphene/WSe$_2$ junction is not gated and possibly also because of an inhomogeneous contact between graphene and WSe$_2$. Also note that in this particular device the WSe$_2$ flake bends over the bottom electrode edge, facilitating the detection of the emission from dark excitons with out-of-plane dipole (see Figs. 2c and 2e in the main text).

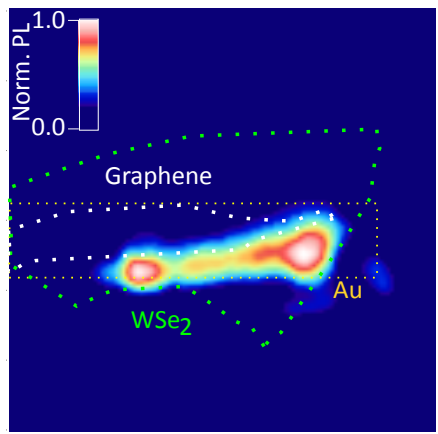

**Supplementary Figure 2 | Photoluminescence map at 300 K (WSe$_2$ sample).** Photoluminescence from monolayer WSe$_2$ at room temperature (excitation $\lambda = 532$ nm, $P_d = 8\times10^3$ W cm$^{-2}$). The PL is quenched on graphene.

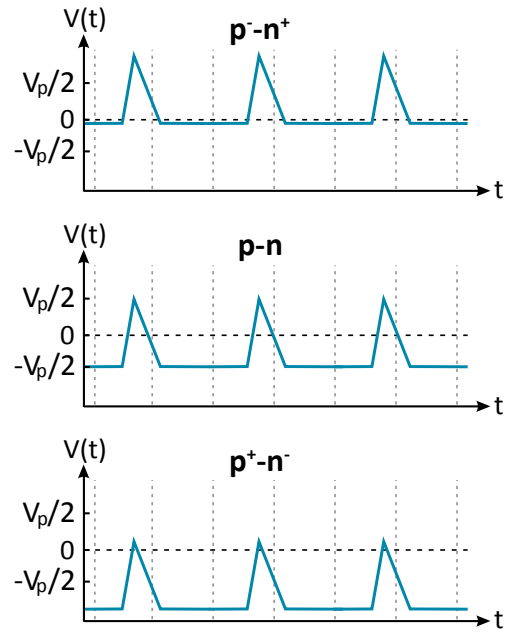

**Supplementary Figure 3 | Schematic pulse diagrams.** Top: electron-rich ($p^-/n^+$); Middle: balanced (p/n); Bottom: hole-rich ($p^+/n^-$) carrier injection.

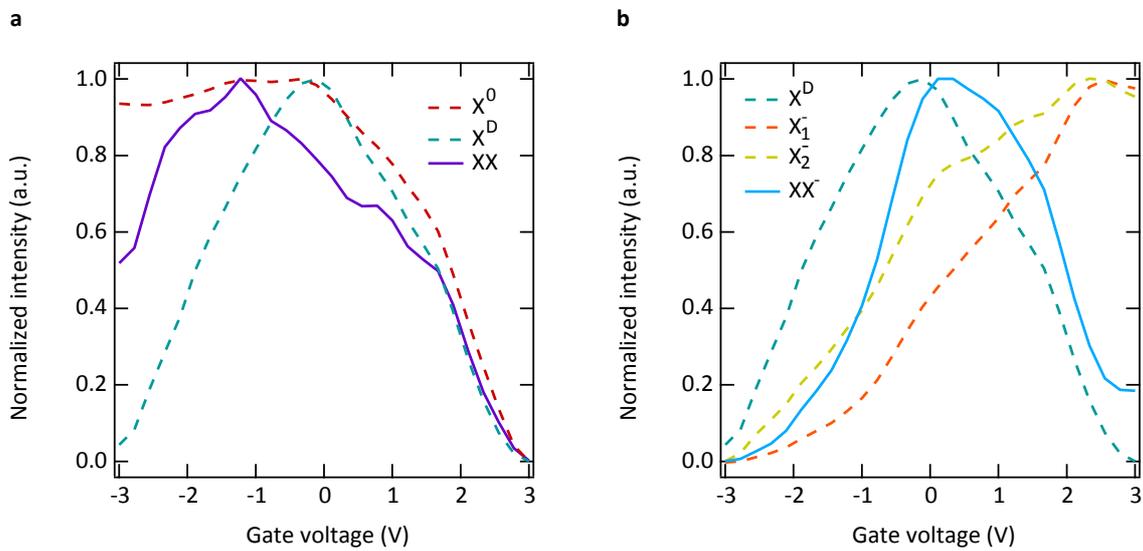

**Supplementary Figure 4 | PL emission from exciton complexes as a function of external electric field in WSe$_2$.** Line-cut of the normalized PL intensity of different exciton species. **a,** Biexcitons (purple) **b,** charged biexcitons (blue) as a function of doping level. The behavior of the different peaks with the back voltage is in good agreement with recently reported results[1–4].

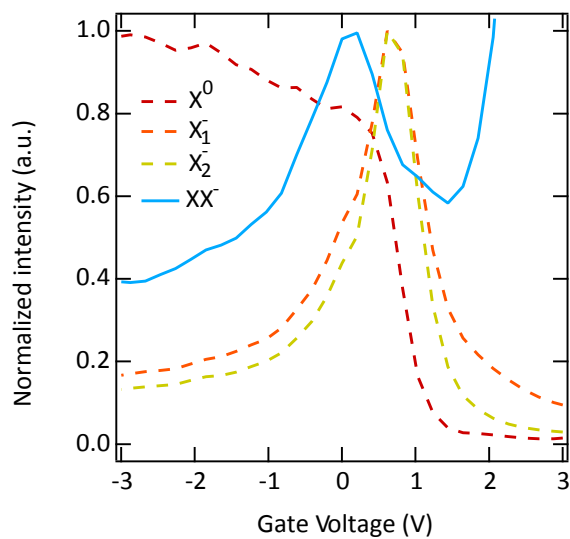

**Supplementary Figure 5 | PL emission from exciton complexes as a function of external electric field in WS$_2$.** Line-cut of the normalized PL intensity of different exciton species as a function of doping level. In the highly n-doped regime ($V_G$ > 2 V), where $X^0$ (red), $X_1^-$ (orange), $X_2^-$ (yellow), $XX^-$ (blue) species vanish, the next charging state of the trion may be observed, as in WSe$_2$[3].

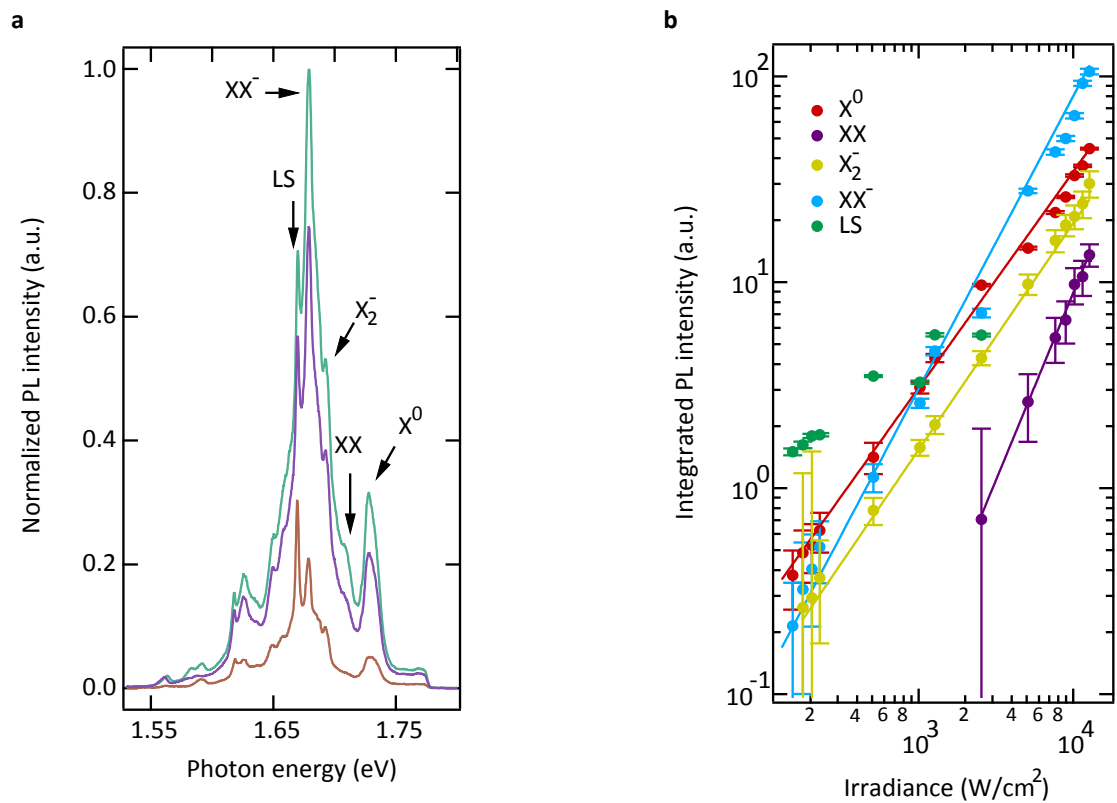

**Supplementary Figure 6 | Power-dependence of the PL intensity in WSe$_2$. a,** PL spectra of a WSe$_2$ monolayer for different excitation intensities ($P_d$ = 2.5x10$^3$ Wcm$^{-2}$ to 1.2x10$^4$ Wcm$^{-2}$). **b,** Logarithmic plot of the integrated PL intensity as a function of the excitation intensity. Circles represent the experimental data, while lines

represent a power law fit, $\propto P^\alpha$, with exponents of $X^0$ ($\alpha$ = 1.01), XX ($\alpha$ = 1.82), $X_2^-$ ($\alpha$ =1.10), $XX^-$ ($\alpha$ = 1.44). The peak LS ($\alpha$ = 0.6), centered at an energy of 1.6696 eV, is most probably due to a localized state, since the dependence on excitation intensity is sublinear[5–10]. The power-dependence measurements are performed without gate voltage (intrinsic regime) at the position marked with a cross in Supplementary Figure 1a.

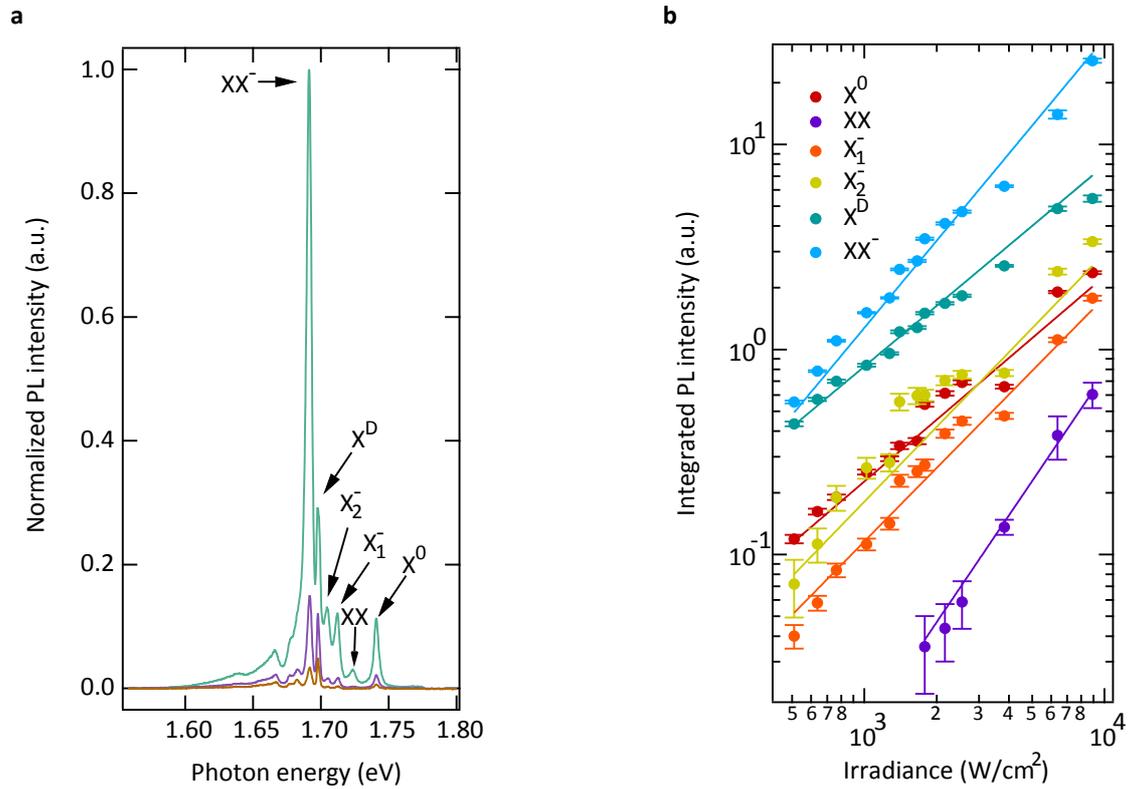

**Supplementary Figure 7 | Power-dependence of the PL intensity in another WSe$_2$ sample. a,** PL spectra of a WSe$_2$ monolayer for different excitation intensities ($P_d$ = 5.0x10$^2$ Wcm$^{-2}$ to 6.3x10$^3$ Wcm$^{-2}$). **b,** Logarithmic plot of the integrated PL intensity as a function of the excitation intensity. Circles represent the experimental data, while lines represent a power law fit, $\propto P^\alpha$, with exponents of $X^0$ ($\alpha$ = 1.01), XX ($\alpha$ = 1.73), $X_1^-$ ($\alpha$ =1.19), $X_2^-$ ($\alpha$ =1.21), $X^D$ ($\alpha$ = 0.99), $XX^-$ ($\alpha$ = 1.41). The power-dependence measurements are performed without gate voltage (intrinsic regime).

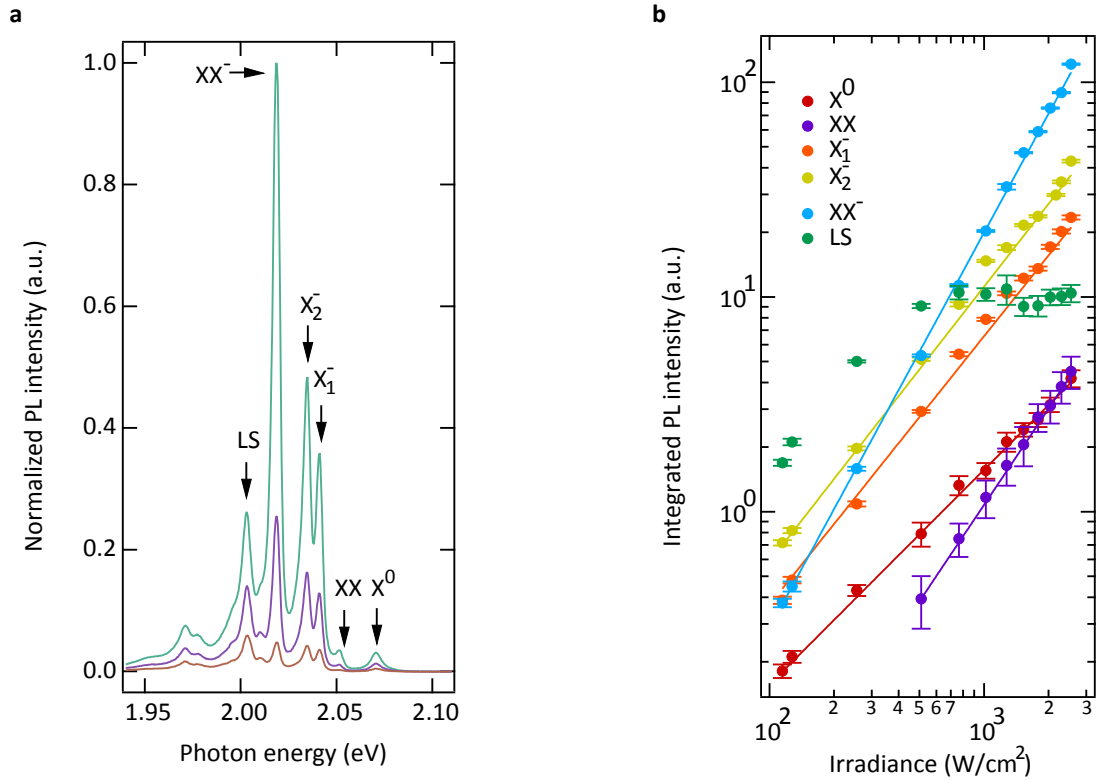

**Supplementary Figure 8 | Power-dependence of the PL intensity in WS$_2$. a,** PL spectra of a WS$_2$ monolayer for different excitation intensities ($P_d$ = 5.1x10$^2$ Wcm$^{-2}$ to 2.5x10$^3$ Wcm$^{-2}$). **b,** Logarithmic plot of the integrated PL intensity as a function of the excitation intensity. Circles represent the experimental data, while lines represent a power law fit, $\propto P^\alpha$, with exponents of X$^0$ ($\alpha$ = 1.01), XX ($\alpha$ = 1.49), X$_1^-$ ($\alpha$ = 1.25), X$_2^-$ ($\alpha$ = 1.28), XX$^-$ ($\alpha$ = 1.84). The peak LS ($\alpha$ = 0.2) at an energy of 2.006 eV, is most probably due to a localized state or phonon-assisted emission[11,12]. The power-dependence measurements are performed without gate voltage (intrinsic regime) on the position marked with a cross in Figure 1b of the main text.

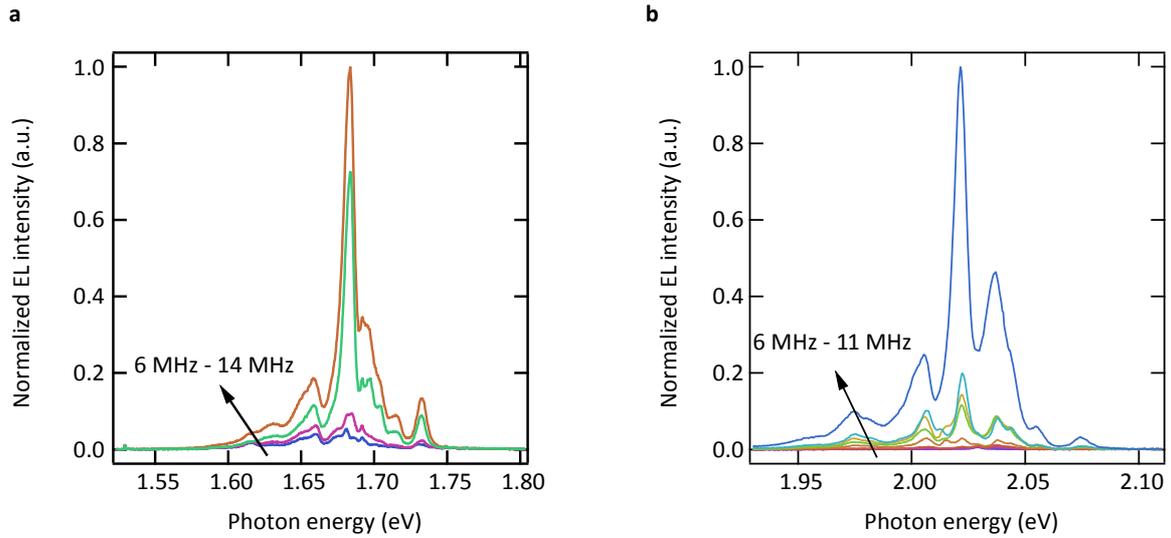

**Supplementary Figure 9 | EL intensity as a function of the pulse frequency. a,** $WSe_2$ EL spectra at frequencies ranging from 6 MHz to 14 MHz. **b,** $WS_2$ EL spectra at frequencies ranging from 6 MHz to 11 MHz. The emission intensity increases with frequency, in agreement with similar measurements performed at room temperature[13]. Furthermore, all the emission peaks broaden at higher frequencies.

|  |  | $X^0$ | XX | $X_1^-$ | $X_2^-$ | $X^D$ | $XX^-$ | $L_1$ |
|---|---|---|---|---|---|---|---|---|
| $WSe_2$ |  |  |  |  |  |  |  |  |
|  | EL | 1 | 1.44 | - | - | 1.03 | 1.75 | - |
|  | PL | 1.01 | 1.82 | 1.19* | 1.1 | 1.08 | 1.44 | 0.6 |
| $WS_2$ |  |  |  |  |  |  |  |  |
|  | EL | 1 | 2.33 | 1.32 | 1.33 | - | 2.31 | - |
|  | PL | 1.00 | 1.49 | 1.25 | 1.28 | - | 1.84 | 0.2 |

**Supplementary Table 1 | α coefficient values obtained from the power-law fits in EL and PL for the different exciton species.** *This value is obtained from measurements performed in a $WSe_2$ sample different from that presented in the main text and shown in Supplementary Figure 7.